# High-energy particle acceleration in the shell of a supernova remnant

F. A. Aharonian[1], A. G. Akhperjanian[2], K.-M. Aye[3], A. R. Bazer-Bachi[4], M. Beilicke[5], W. Benbow[1], D. Berge[1], P. Berghaus[6]*, K. Bernlöhr[1,7], O. Bolz[1], C. Boisson[8], C. Borgmeier[7], F. Breitling[7], A. M. Brown[3], J. Bussons Gordo[9], P. M. Chadwick[3], V. R. Chitnis[10,20]*, L.-M. Chounet[11], R. Cornils[5], L. Costamante[1,20], B. Degrange[11], A. Djannati-Ataï[6], L. O'C. Drury[12], T. Ergin[7], P. Espigat[6], F. Feinstein[9], P. Fleury[11], G. Fontaine[11], S. Funk[1], Y. A. Gallant[9], B. Giebels[11], S. Gillessen[1], P. Goret[13], J. Guy[10], C. Hadjichristidis[3], M. Hauser[14], G. Heinzelmann[5], G. Henri[15], G. Hermann[1], J. A. Hinton[1], W. Hofmann[1], M. Holleran[16], D. Horns[1], O. C. de Jager[16], I. Jung[1,14]*, B. Khélifi[1], Nu. Komin[7], A. Konopelko[1,7], I. J. Latham[3], R. Le Gallou[3], M. Lemoine[11], A. Lemière[6], N. Leroy[11], T. Lohse[7], A. Marcowith[4], C. Masterson[1,20], T. J. L. McComb[3], M. de Naurois[10], S. J. Nolan[3], A. Noutsos[3], K. J. Orford[3], J. L. Osborne[3], M. Ouchrif[10,20], M. Panter[1], G. Pelletier[15], S. Pita[6], M. Pohl[17]*, G. Pühlhofer[1,14], M. Punch[6], B. C. Raubenheimer[16], M. Raue[5], J. Raux[10], S. M. Rayner[3], I. Redondo[11,20]*, A. Reimer[17], O. Reimer[17], J. Ripken[5], M. Rivoal[10], L. Rob[18], L. Rolland[10], G. Rowell[1], V. Sahakian[2], L. Saugé[15], S. Schlenker[7], R. Schlickeiser[17], C. Schuster[17], U. Schwanke[7], M. Siewert[17], H. Sol[8], R. Steenkamp[19], C. Stegmann[7], J.-P. Tavernet[10], C. G. Théoret[6], M. Tluczykont[11,20], D. J. van der Walt[16], G. Vasileiadis[9], P. Vincent[10], B. Visser[16], H. J. Völk[1] & S. J. Wagner[14]

[1]*Max-Planck-Institut für Kernphysik, PO Box 103980, D 69029 Heidelberg, Germany*

[2]*Yerevan Physics Institute, 2 Alikhanian Brothers Street, 375036 Yerevan, Armenia*

[3]*Department of Physics, University of Durham, South Road, Durham DH1 3LE, UK*

[4]*Centre d'Etude Spatiale des Rayonnements, CNRS/UPS, 9 av. du Colonel Roche, BP 4346, F-31029 Toulouse Cedex 4, France*

[5] *Institut für Experimentalphysik, Universität Hamburg, Luruper Chaussee 149, D 22761 Hamburg, Germany*

[6]*Physique Corpusculaire et Cosmologie, IN2P3/CNRS, Collège de France, 11 Place Marcelin Berthelot, F-75231 Paris Cedex 05, France*

[7]*Institut für Physik, Humboldt-Universität zu Berlin, Newtonstrasse 15, D 12489 Berlin, Germany*

[8]*LUTH, UMR 8102 du CNRS, Observatoire de Paris, Section de Meudon, F-92195 Meudon Cedex, France*

[9]*Groupe d'Astroparticules de Montpellier, IN2P3/CNRS, Université Montpellier II, CC85, Place Eugène Bataillon, F-34095 Montpellier Cedex 5, France*

[10]*Laboratoire de Physique Nucléaire et de Hautes Energies, IN2P3/CNRS, Universités Paris VI & VII, 4 Place Jussieu, F-75231 Paris Cedex 05, France*

[11]*Laboratoire Leprince-Ringuet, IN2P3/CNRS, Ecole Polytechnique, F-91128 Palaiseau, France*

[12]*Dublin Institute for Advanced Studies, 5 Merrion Square, Dublin 2, Ireland*

[13]*Service d'Astrophysique, DAPNIA/DSM/CEA, CE Saclay, F-91191 Gif-sur-Yvette, France*

[14]*Landessternwarte, Königstuhl, D 69117 Heidelberg, Germany*

[15]*Laboratoire d'Astrophysique de Grenoble, INSU/CNRS, Université Joseph Fourier, BP 53, F-38041 Grenoble Cedex 9, France*

[16]*Unit for Space Physics, North-West University, Potchefstroom 2520, South Africa*

[17]*Institut für Theoretische Physik, Lehrstuhl IV: Weltraum und Astrophysik, Ruhr-Universität Bochum, D 44780 Bochum, Germany*

[18]*Institute of Particle and Nuclear Physics, Charles University, V Holesovickach 2, 180 00 Prague 8, Czech Republic*

[19]*University of Namibia, Private Bag 13301, Windhoek, Namibia*

[20]*European Associated Laboratory for Gamma-Ray Astronomy, jointly supported by CNRS and MPG*

*Present addresses: Université Libre de Bruxelles, Faculté des Sciences, Campus de la Plaine, CP230, Boulevard du Triomphe, 1050 Bruxelles, Belgium (P.B.); Tata Institute of Fundamental Research, Homi Bhabha Road, Mumbai 400 005, India (V.R.C.); Washington University, Department of Physics, 1 Brookings Drive, CB 1105, St Louis, Missouri 63130, USA (I.J.); Department of Physics and Astronomy, Iowa State University, Ames, Iowa 50011-3160, USA (M.P.); Department of Physics and Astronomy, University of Sheffield, The Hicks Building, Hounsfield Road, Sheffield S3 7RH, UK (I.R.)

**A significant fraction of the energy density of the interstellar medium is in the form of high-energy charged particles (cosmic rays)[1]. The origin of these particles remains uncertain. Although it is generally accepted that the only sources capable of supplying the energy required to accelerate the bulk of Galactic cosmic rays are supernova explosions, and even though the mechanism of particle acceleration in expanding supernova remnant (SNR) shocks is thought to be well understood theoretically[2,3], unequivocal evidence for the production of high-energy particles in supernova shells has proven remarkably hard to find. Here we report on observations of the SNR RX J1713.7−3946 (G347.3−0.5), which was discovered by ROSAT[4] in the X-ray spectrum and later claimed as a source of high-energy γ-rays[5,6] of TeV energies (1 TeV=$10^{12}$ eV). We present a TeV γ-ray image of the SNR: the spatially resolved remnant has a shell morphology similar to that seen in X-rays, which demonstrates that very-high-energy particles are accelerated there. The energy spectrum indicates efficient acceleration of charged particles to energies beyond 100 TeV, consistent with current ideas of particle acceleration in young SNR shocks.**

RX J1713.7−3946, together with several other southern hemisphere SNRs, is a prime target for observations with the High Energy Stereoscopic System (H.E.S.S.), a new system of four imaging atmospheric Cherenkov telescopes located in the Khomas Highland of Namibia. H.E.S.S.[7,8] (we note that V. F. Hess discovered cosmic rays) exploits the most effective detection technique for very-high-energy γ-rays, namely, the imaging of Cherenkov light from air showers. This technique, which was pioneered by the Whipple collaboration[9], makes use of the fact that whenever a high-energy γ-ray hits the Earth's atmosphere it is absorbed and initiates a cascade of interactions with air atoms, leading to the formation of a shower of secondary charged particles. Those travelling faster than the local speed of light in air emit Cherenkov radiation, which results in a brief flash of blue Cherenkov light detectable at ground level. By using a telescope with sufficient mirror area to collect enough of the faint light signal, and a fast camera with fine pixelation, one can image the shower and reconstruct from this image the direction and energy of the primary γ-ray.

Combined with the approach of stereoscopic imaging of the cascade using a system of telescopes, as pioneered by the HEGRA collaboration[10], this yields a very powerful technique for imaging and obtaining energy spectra of astronomical sources at TeV energies.

The H.E.S.S. experiment is such a stereoscopic system that consists of four 13-m-diameter telescopes[11] spaced at the corners of a square of side 120 m, each equipped with a 960-phototube camera[12] covering a large field of view of diameter 5°. Construction of the telescope system started in 2001; the full array was completed in December 2003 with the commissioning of the fourth telescope. HESS has an angular resolution of a few arc minutes, an effective energy range from 100 GeV to 10 TeV with energy resolution of 15–20% and a flux sensitivity approaching $10^{-13}$ erg cm$^{-2}$ s$^{-1}$. These characteristics, together with its southern hemisphere location, make HESS ideally suited for spectroscopic and morphological studies of Galactic plane sources such as RX J1713.7−3946, which is now the first SNR shell to be confirmed as a TeV source. TeV emission has also been reported from the remnant of SN 1006[13], a result which our observations have not yet allowed us to confirm with H.E.S.S.[14], and from Cassiopeia A[15,16] (a classical core-collapse SNR and the weakest TeV source yet reported) whose northern location makes it inaccessible to H.E.S.S..

Here we present results from observations of RX J1713.7−3946 performed between May and August 2003 during two phases of the construction and commissioning of HESS. In the first phase, two telescopes were operated independently, with stereoscopic event selection done offline using GPS time stamps to identify coincident events. During the second phase, also using two telescopes, coincident events were selected in hardware using an array level trigger. The total on-source observation time was 26 h; after run selection and dead time correction a data set corresponding to 18.1 live hours was used in this analysis. At the trigger level (for observation altitude angles of 60–75°), the energy thresholds for the two configurations were 250 GeV (without the array level trigger) and 150 GeV (with the array level trigger). In this analysis, hard cuts were used to select only well reconstructed showers. This primarily served to drastically reduce the number of background cosmic-ray events, but it also homogenized these data taken with the two different configurations and improved the angular resolution. Thereby systematic errors are greatly reduced at the expense of a higher energy threshold (~800 GeV for the combined data set).

Figure 1 shows the resulting count map centred on RX J1713.7−3946. The SNR stands out from the residual charged cosmic-ray background with a significance of 20 standard deviations. The overall shell structure is clearly visible and coincides closely with that seen in X-rays (see Fig. 2). To our knowledge, this is the first case where the identification of an active, that is accelerating, celestial γ-ray source (as opposed to a passive cloud or density enhancement, merely penetrated by energetic particles which are accelerated elsewhere) can be based, not just on a positional coincidence, but on the image morphology. The overall flux above 1 TeV is (1.46 ± 0.17 (statistical) ± 0.37 (systematic)) × $10^{-7}$ photons m$^{-2}$ s$^{-1}$, which corresponds to 66% of the Crab nebula flux as measured by H.E.S.S.. More elaborate analyses of these data using different background models (needed to determine the spectrum) and independent (and different) analysis chains confirm the results presented here.

The energy spectrum of the whole remnant is shown in Fig. 3. These data are well described by a power law (see Fig. 3 legend) with a photon index $\Gamma = 2.19 \pm 0.09 \pm 0.15$, as compared to the photon index of $2.84 \pm 0.15 \pm 0.20$ reported by the CANGAROO-II collaboration for the northwest part of the SNR[6]. The integral energy flux between 1 TeV and 10 TeV is estimated to be $3.5 \times 10^{-11}$ erg cm$^{-2}$ s$^{-1}$, which is an order of magnitude smaller than the non-thermal X-ray flux. More data will be taken with the full H.E.S.S. array in 2004. The increased sensitivity (four instead of two telescopes) will enable spatially resolved spectral studies.

RX J1713.7−3946 (situated in the Galactic plane, in the constellation Scorpius) is one of the brighter Galactic X-ray SNRs[17], with a flux density of a few times $10^{-10}$ erg cm$^{-2}$ s$^{-1}$. In X-rays, it exhibits typical shell morphology, but remarkably the X-ray spectrum is completely dominated by a non-thermal continuum with no detectable line emission. The most plausible origin of these X-rays is the synchrotron radiation of 100-TeV electrons[18,19]. However, because alternative explanations are not absolutely ruled out[20], only the detection of TeV emission from this SNR will provide unambiguous evidence for acceleration of particles to multi-TeV energies. Another important point is that the TeV signal may contain a component due to accelerated protons interacting with the ambient gas, as has been predicted theoretically[2,3]. The contribution of this component should be significantly enhanced when the supernova shell overtakes nearby dense molecular clouds[21], as seems to be the case for this object. The CO data[22] suggest that a cloud is interacting with the northwestern part of the SNR, where a striking spatial coincidence between the CO density peaks and the regions of peak X-ray emission is seen. The X-ray data[23] also indicate significant absorption column densities in the western part of the remnant, at values about twice those to the east. These indications fit qualitatively with the γ-ray image presented here, where TeV emission is seen from the whole SNR shell but with an increased flux from the northwestern side.

Given the recent estimates[22–24] of the distance to the source of 1 kpc, if a significant part of the TeV flux were to be formed by interactions of cosmic-ray nuclei with gas atoms in the cloud with density $n$ exceeding 100 cm$^{-3}$, the energetics implied by the γ-ray flux and the spectrum would be a few times $10^{49} n^{-1}$ erg between 10 and 100 TeV. This is consistent with the picture of an SNR origin of Galactic cosmic rays involving about 10% efficiency for conversion of the mechanical energy of the explosion into non-thermal particles, and a production spectrum in the SNR which is approximately an $E^{-2}$ power law from several GeV to about one PeV. Moreover, the γ-ray morphology is qualitatively what one would expect from particles accelerated at the shock interacting and radiating in the compressed post-shock region. The extension of the γ-ray spectrum up to energies of 10 TeV (see Fig. 3) requires an extremely effective accelerator boosting particles up to energies of at least 100 TeV. RX J1713.7−3946 is a complex object interacting with molecular clouds of different densities where the TeV emission might emerge from various processes. Without doubt there will be a contribution from energetic electrons through the inverse Compton process, especially from low-density regions (as in the eastern part) of the SNR. At the elevated densities likely to exist in the northwestern rim, $\pi^0$-decays following proton–proton interactions, but also non-thermal Bremsstrahlung of electrons, could make significant contributions. Although disentangling the relative contributions of the various processes is difficult, it should be possible through spatially resolved multi-wavelength studies—which will be undertaken using the full H.E.S.S. array.

The multi-TeV image of a shell-type SNR presented here constitutes a significant step forward towards a solution of the long-standing puzzle of the origin of Galactic cosmic rays, as well as demonstrating an astronomical imaging technique operating at photon energies some 12 decades higher than those of visible light.

**Acknowledgements** The support of the Namibian authorities and of the University of Namibia in facilitating the construction and operation of HESS is acknowledged. We also thank the following for support: the German Ministry for Education and Research (BMBF), the Max Planck Society, the French Ministry for Research, the CNRS-IN2P3 and the Astroparticle Interdisciplinary Programme of the CNRS, the UK Particle Physics and Astronomy Research Council

(PPARC), the IPNP of Charles University, the South African Department of Science and Technology and National Research Foundation, and the University of Namibia. The European Associated Laboratory for Gamma-Ray Astronomy is jointly supported by CNRS and MPG. We appreciate the work of the technical support staff in Berlin, Durham, Hamburg, Heidelberg, Palaiseau, Paris, Saclay and Namibia in the construction and operation of the equipment. We also thank Y. Uchiyama for supplying the ASCA X-ray data shown in Fig. 1.

**Correspondence** and requests for materials should be addressed to D.B. (David.Berge@mpi-hd.mpg.de).

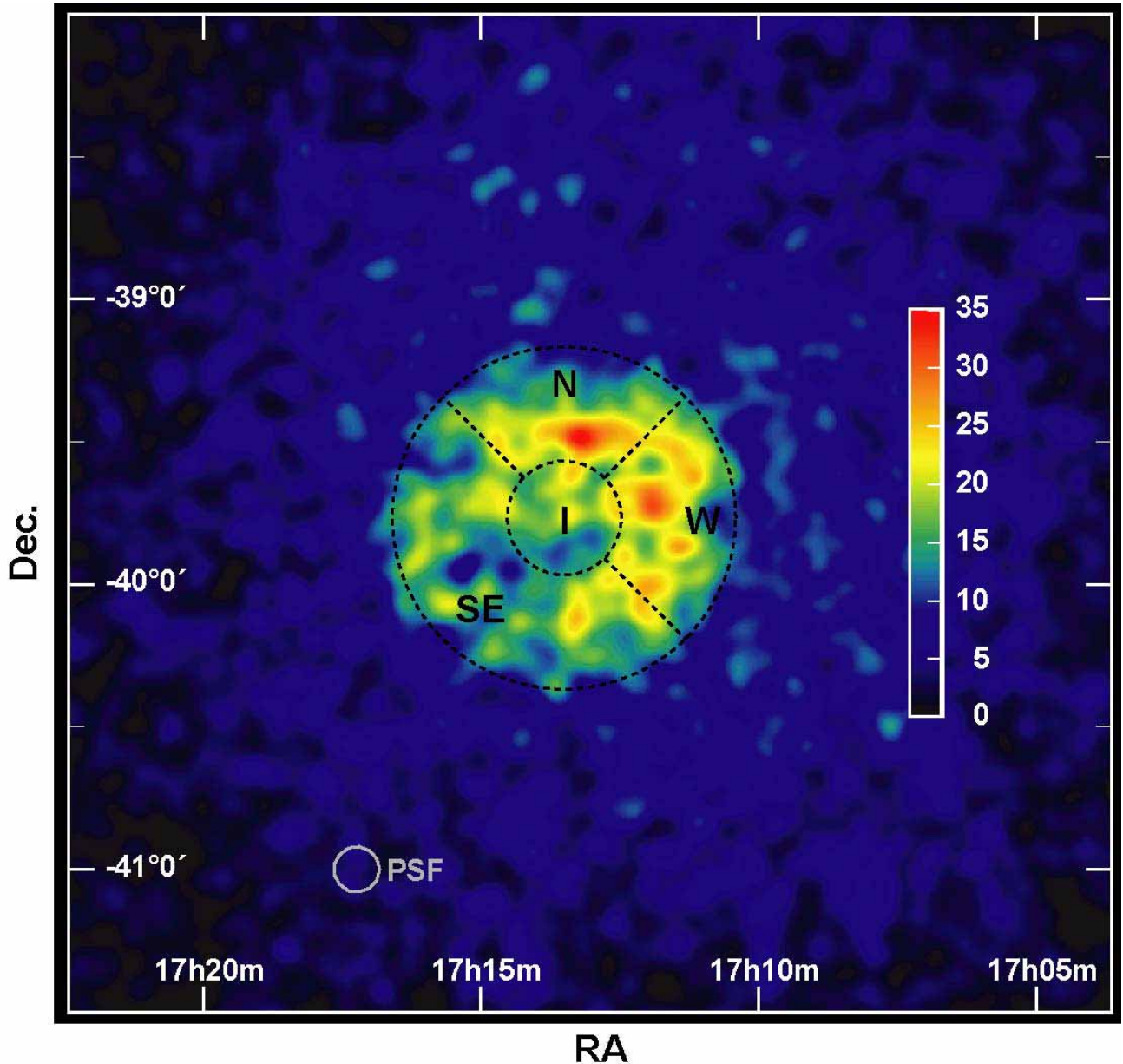

**Figure 1** Wide field of view (3.5° × 3.5°) around RX J1713.7−3946. Integration regions for flux estimates are shown. The flux above 1 TeV from the northern (N) rim is $(3.0 \pm 0.6) \times 10^{-8}$ photons m$^{-2}$ s$^{-1}$, from the western (W) rim $(4.1 \pm 0.8) \times 10^{-8}$ photons m$^{-2}$ s$^{-1}$, from the southeastern (SE) rim $(5.9 \pm 1.0) \times 10^{-8}$ photons m$^{-2}$ s$^{-1}$ and the flux from the interior (I) $(1.7 \pm 0.6) \times 10^{-8}$ photons m$^{-2}$ s$^{-1}$. The mean γ-ray brightnesses per unit solid angle (that is, the flux values normalized to the area) of the regions are in the ratio 1:1.4:1:1.2 (N:W:SE:I). These values might contradict the visual impression that the northwestern shell of the SNR is brighter. However, statistics of the data sample are limited, and the different areas were chosen for geometric reasons. Looking at the image, we see for example that dim regions are included in the seemingly brighter northern and western area, whereas the interior might gain from leakages from the northwestern shell. More detailed spatially resolved flux studies will have to await the advent of new data taken with the full H.E.S.S. array with increased sensitivity. The 70% containment radius of the γ-ray point-spread function (PSF) for this data set with an energy threshold of 800 GeV is illustrated in the bottom left-hand corner (structures that are smaller than this circle should not be considered as real). The image is smoothed with a Gaussian of standard deviation 3 arcmin (matched to the angular resolution of the instrument for this particular data set). The linear color scale is in units of counts. Note that the efficiency of the camera falls off towards the edge of the field of view.

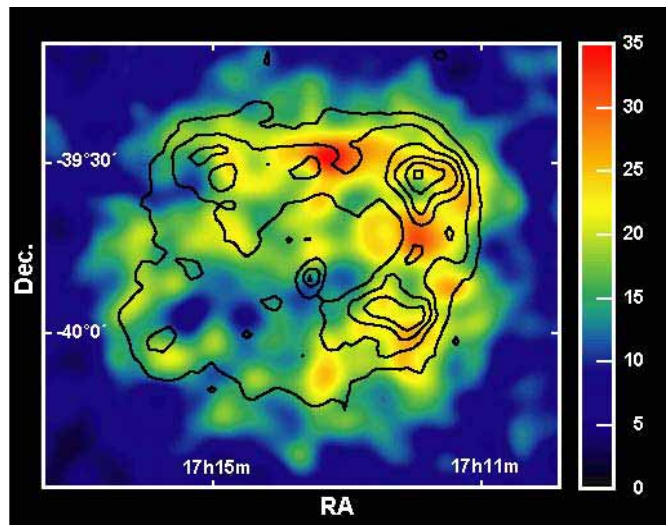

**Figure 2** γ-ray image of the SNR RX J1713.7−3946 obtained with the H.E.S.S. telescopes. Hard cuts were applied to select well-reconstructed γ-like events above 800 GeV. The map is smoothed as in Fig. 1, having the same scale in units of counts. We note that no background subtraction or camera-efficiency corrections have been applied. This demonstrates that the structures seen are not artefacts of the analysis but real and visible in the raw post-cuts data (the background in the field of view is at a level of about five counts, and the efficiency across the SNR changes by less than 10%). This image, obtained with a partial array during construction, demonstrates the ability of H.E.S.S. to map extended objects. The superimposed (linearly spaced) contours show the X-ray surface brightness as seen by ASCA in the 1–3 keV range for comparison[25]. Note that the angular resolution of ASCA is comparable to that of H.E.S.S. which enables direct comparison of the two images. RA, right ascension; dec., declination.

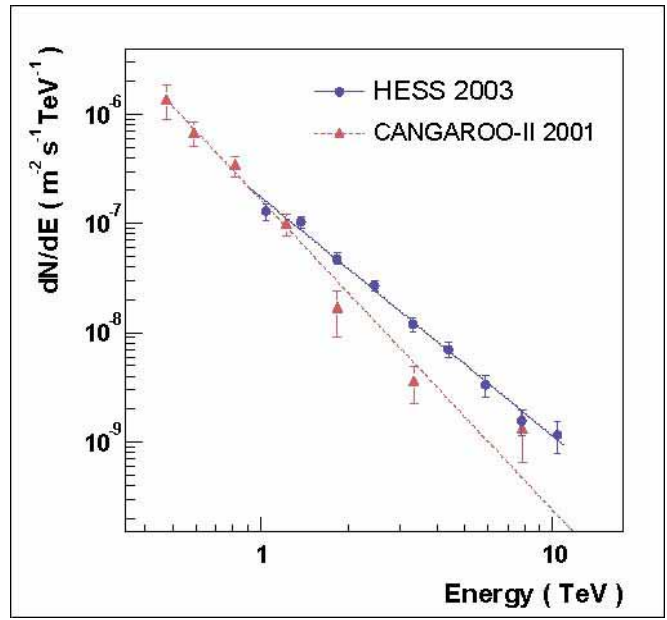

**Figure 3** γ-ray energy spectrum of RX J1713.7−3946 as measured with the HESS telescopes. These data (blue filled circles) can be described by a power law, $dN/dE \propto E^{-\Gamma}$; the best fit result (blue solid line) gives $\Gamma = 2.19 \pm 0.09$ (statistical) $\pm 0.15$ (systematic) with $\chi^2 = 5.9$ with 7 degrees of freedom. The integral flux above 1 TeV is found to be $(1.46 \pm 0.17$ (statistical) $\pm 0.37$ (systematic)$) \times 10^{-7}$ photons m$^{-2}$ s$^{-1}$. There is clearly no evidence for a cut-off in the data, but if one nevertheless attempts to fit an exponentially cut-off power law of the form $dN/dE \propto E^{-\Gamma_c} e^{-E/E_c}$, the minimum acceptable value of $E_c$ is 4 TeV with a very hard photon index of $\Gamma_c = 1.5$. The spectral data points reported by the CANGAROO-II collaboration[6] for the northwestern part of the remnant are shown for comparison as red triangles, and the best fit result as a dashed red line. The fact that CANGAROO-II reported a spectrum only for a part of the SNR prohibits at this stage a definite statement about the compatibility of the two measurements. Error bars, $1\sigma$ statistical errors.